\documentclass[showpacs,prl,superscriptaddress,twocolumn]{revtex4}

\usepackage{graphicx}
\usepackage{dcolumn}
\usepackage{bm}

\begin{document}

\title{Surface-edge state and half quantized Hall conductance in topological
insulators}
\author{Rui-Lin Chu}
\affiliation{Department of Physics and Center of Computational and Theoretical Physics,
The University of Hong Kong, Pokfulam Road, Hong Kong}
\author{Junren Shi}
\affiliation{International Center for Quantum Materials, Peking University, Beijing
100871, China}
\author{Shun-Qing Shen}
\affiliation{Department of Physics and Center of Computational and Theoretical Physics,
The University of Hong Kong, Pokfulam Road, Hong Kong}
\date{\today }

\begin{abstract}
We investigate the transport properties of the surface states of a
three-dimensional topological insulator in the presence of a spin-splitting
Zeeman field. We propose a picture that the chiral edge state forms on
the surface, and is split into two halves that are spatially
separated, each carrying one half of the conductance quantum ($e^{2}/h$).
This picture is confirmed by numerical simulation in a four-terminal setup.
It is demonstrated that the difference between the clockwise and counterclockwise transmission coefficients of the two neighboring terminals is 
approximately one half, which suggests that the half quantized Hall
conductance can be manifested in an appropriate experimental setup.
\end{abstract}

\maketitle

Topological insulators are insulating in the bulk, but have metallic surface
states possessing an odd number of Dirac cones of massless fermions~\cite%
{Moore-10Nature,Qi-10PhysTo,Hasan-10RMP}. The band structure and the quantum
spin texture of these surfaces states have been well established
theoretically and experimentally~\cite%
{Fu-07prl,Moore-07prb,Hsieh-08Nature,Zhang-09NP,Xia-09NP,Chen-09xxx}. In the
presence of a spin-splitting Zeeman field, which could be induced by
magnetically doping the samples or putting the samples in the proximity of
ferromagnetic materials, the Dirac fermions will gain a mass and the
spectrum opens a gap~\cite{Yokoyama-10PRL,Garate-10PRL}. When the Fermi
level is located within the energy gap, it was proposed that the Hall
conductance of the surface states will be one half of the conductance
quantum $e^{2}/h$~\cite{Redlich-84prd,Jackiw-84prd,Qi-08prb}. Based on this,
Qi et al. proposed the unconventional magneto-electric effect, which is
regarded as one of the characteristic features of the topological insulators~%
\cite{Qi-08prb,Qi-09Science,Essin-09prl}.

On the other hand, it is not clear whether or not the half quantization of
the Hall conductance can be directly observed in the transport measurement.
In the usual quantum Hall system, the current-carrying chiral edge states
are responsible for the integer quantized Hall conductance measured in the
transport experiment.\cite{Halperin-82prb,MacDonald-84prb} It is not
immediately clear whether or not the similar chiral edge state will form on
the closed surface of a topological insulator, and how the quantized nature
of the edge states can be reconciled with the prediction of the half
quantization of the Hall conductance.\cite%
{Redlich-84prd,Jackiw-84prd,Qi-08prb}

To get a definite answer to these questions, we investigate the
multi-terminal transport properties of a 3D topological insulator in the
presence of a uniform spin-splitting Zeeman field. We propose that the the
closed surface of the topological insulator will form two insulating domains
of the different topologies (i.e., positive vs. negative gaps), separated by
a gapless metallic belt. A chiral edge state will form and is concentrated
around the boundaries between the insulating domains and the metallic belt.
Effectively, the chiral edge state is split into two halves, each of which
is circulating around the boundary of one of the domains and carrying one
half of the conductance quantum ($e^{2}/h$). Such a picture reconciles the
apparent conflict between the half quantization and the index theorem. It
also suggests that by attaching the electrodes to the boundary of one of
these domains, it is possible to directly measure the half quantized Hall
conductance. The picture as well as its prediction are verified by the
numerical simulation based non-equilibrium Green's function method.

\begin{figure}[tph]
\centering \includegraphics[height=50mm]{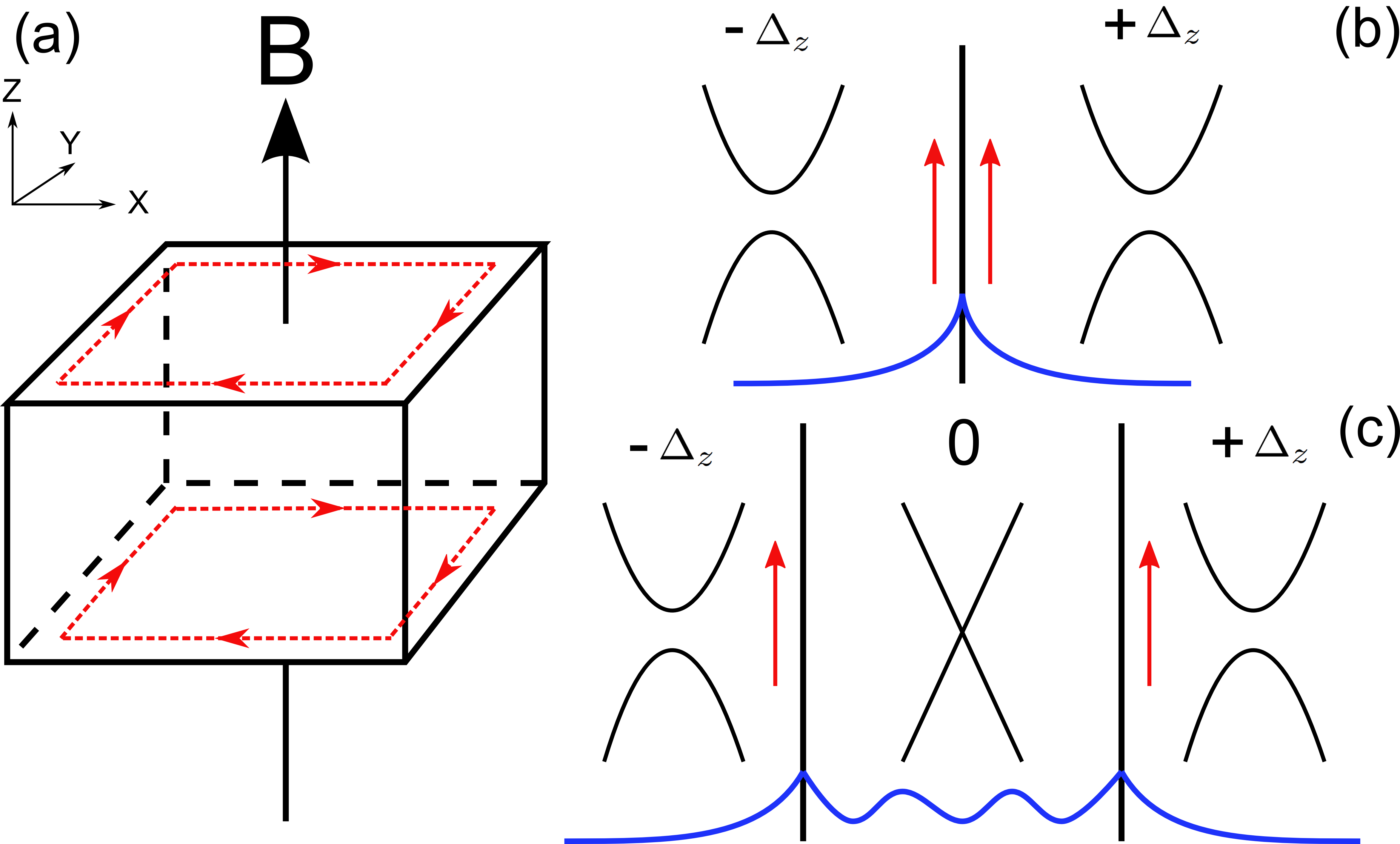}
\caption{(Color online) (a) Schematic of a 3D TI in a weak Zeeman field, and
the formation of chiral current on the top and bottom surface boundaries.
(b) A chiral edge state will form around the domain wall between the 2D
Dirac fermions with positive and negative masses, and the wave function is
illustrated. The arrow indicates the flow of edge current. (c) When the
sharp domain wall evolves to finite-width metallic band, the edge mode is
effectively split to two halves concentrated around the two boundaries.}
\end{figure}

To illustrate the basic physics, we consider a 3D topological insulator of
the cubic shape. A Zeeman field is applied along the $z$-direction, as shown
in Fig.~1(a). Because the bulk of the system is insulating, it is
effectively a closed 2D surface with six facets. The effective Hamiltonian
of the Dirac fermions for the surface state can be written as~\cite%
{Shan-10njp,Lu-10prb}: 
\begin{equation}
H_{\mathrm{eff}}(\bm k)=v\left( \bm{k}\times \bm{\sigma}\right) \cdot \bm{n}%
-g_{\parallel }\mu _{B}h_{\parallel }\sigma _{\parallel }-g_{\perp }\mu _{B}%
\bm{h}_{\perp }\cdot \bm{\sigma}_{\perp },  \label{effectiveH}
\end{equation}%
where $\bm{n}$ denotes the normal vector of the surface, $\bm{\sigma}\equiv
\{\sigma _{x},\sigma _{y},\sigma _{z}\}$ are the Pauli matrices, $%
h_{\parallel }$ ($\sigma _{\parallel }$) and $\bm h_{\perp }$ ($\bm\sigma
_{\perp }$) are the Zeeman field (Pauli matrix) components parallel and
perpendicular to the normal vector, respectively, and $g_{\parallel }$ and $%
g_{\perp }$ are the corresponding spin $g$-factor. Note that the surface
state has the anisotropic spin $g$-factor due to the strong spin-orbit
coupling of the bulk band: $g_{\parallel }$ is the same as the $g$-factor of
the bulk material, and $g_{\perp }$ is renormalized by bulk band parameters
and is usually strongly suppressed~\cite{Shan-10njp,Lu-10prb}. The different
facets of the surface have the different effective Hamiltonians respective
to the different normal vectors $\bm n$. For the top and the bottom facets,
the effective Hamiltonian can be written as $H_{\mathrm{eff}}=\pm
v(k_{x}\sigma _{y}-k_{y}\sigma _{x})+\Delta _{z}\sigma _{z}$, where $+$ and $%
-$ are for the top and bottom surfaces, respectively, and $\Delta _{z}\equiv
-g_{\parallel }\mu _{B}h$. The spectrum will open a gap on these facets, and
the Dirac fermions gain a mass $\pm \Delta _{z}$. On the other hand, the
effective Hamiltonians for the side facets can be written as the forms like $%
H_{\mathrm{eff}}=v[(k_{x}+\Delta k)\sigma _{z}-k_{x}\sigma _{x}]$, where $%
\Delta k\equiv g_{\perp }\mu _{B}h$. In this case, the Zeeman field has no
effect except shifting the Dirac point. When the fermi level is located in
the gap of the top and bottom surface, the system becomes effectively two
insulating domains separated by a conducting belt with massless Dirac
fermions.

The two insulating domains belong to the different topological classes due
to the fact that the normal vectors for the top and the bottom surfaces are
in the opposite directions. We consider the limiting case that the sample
has the vanishing thickness along the $z$-direction (but ignore the coupling
between the top and bottom surfaces). The system is locally equivalent to a
domain-wall structure across which the Dirac fermion masses changes from
positive to negative, as shown in Fig.~1(b). The change of the band topology
necessitates the close of the energy gap around the domain wall, giving rise
to the gapless edge mode. By directly solving the wave equation and matching
the wave functions at the boundary, it is easy to obtain that the wave
function of the edge mode has the form $\Psi (x,y)=\sqrt{{\Delta _{z}}/{4\pi
v}}(1,1)^{T}\exp [-({\Delta _{z}}/{v})|x|+\mathrm{i}k_{y}y]$, and the
dispersion is linear in the momentum: $E=-vk_{y}$. The edge mode will form
an ideal one-dimensional chiral edge channel carrying a quantized
conductance $e^{2}/h$. The channel cannot be backscattered and is robust
against the disorder. Transport measurement will yield a quantized Hall
conductance if attaching electrodes to the edge of the sample.

When the thickness of the sample along the $z$-direction is finite, the two
insulating domains are separated by a finite metallic region of massless
Dirac fermion, equivalent to the configuration shown in Fig. 1(c). In this
case, the electrons in the metallic region will in general form sub-bands
due to the confinement by the insulating domains. The normal sub-band has
dispersion $E_{n}=\pm v\sqrt{k_{y}^{2}+k_{n}^{2}}$, where $k_{n}\neq 0$ is
one of the discrete momentums of the eigen-modes along the confinement
direction. Beside these, there is a chiral solution with the dispersion $%
E=-vk_{y}$, corresponding to $k_{n}=0$ but with only the left-going mode.
The chirality is also shown in the penetration amplitude of the normal
sub-band wave function into the insulating domains: the penetration
amplitude maximizes when $E\sim -vk_{y}$, and decreases when moving away
from it. All sub-bands crossing the Fermi energy will contribute to the
transport properties, and form a novel chiral edge state as a result of
collective combination (not a single mode). The chirality will give rise to
a finite Hall conductance.

To see what the transport measurement will yield for such a system, we
employ the direct numerical simulation. We investigate an isotropic 3D
topological insulator with the tight-binding Hamiltonian on a cubic lattice: 
\begin{equation}
\mathcal{H}=\sum_{i}c_{i}^{\dag }\mathcal{M}_{0}c_{i}+\sum_{i,\alpha
=x,y,x}\left( c_{i}^{\dag }\mathcal{T}_{\alpha }c_{i+\alpha }+c_{i+\alpha
}^{\dag }\mathcal{T}_{\alpha }c_{i}\right)
\end{equation}%
where $\mathcal{T}_{a}=B\sigma _{z}\otimes \sigma _{0}-i\frac{A}{2}\sigma
_{x}\otimes \sigma _{\alpha }$ and $M_{0}=(M-6B)\sigma _{z}\otimes \sigma
_{0}+\Delta _{z}\sigma _{0}\otimes \sigma _{z}$. The lattice space is taken
to be unity. Near the $\Gamma $ point in the $k$ space, it is reduced to a
Dirac-like model, which can be derived from either the theory of invariants~%
\cite{Zhang-09NP}, or the 8-band extended Kane model near the inversion of
the $\Gamma _{6}^{-}$ and $\Gamma _{8}^{+}$ bands. The Dirac model for a
topological insulator has been discussed by several authors~\cite%
{Qi-08prb,Murakami}. It was shown that the model yield the effective surface
Hamiltonian Eq.~\ref{effectiveH} with $v=A$~\cite{Shan-10njp,Lu-10prb}. We
first employ the Green function technique to calculate the local density of
states (LDOS) for the surface states, in particular the spatial distribution
near the boundary, to demonstrate the formation of the edge state. Then we
calculate the Hall conductance for a four-terminal setup, which directly
reflects the output of the transport measurement.

\begin{figure}[tph]
\centering \includegraphics[height=50mm]{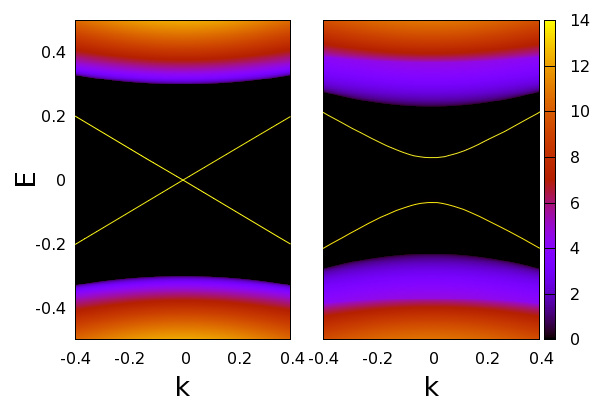}
\caption{(Color online) Local density of state on an infinite xy surface of
a semi-infinite 3D system. (left) gapless single Dirac cone of the surface
state; (right) gap opening by application of a Zeeman splitting term. The
model parameters are $A=0.5$, $B=0.25$, $M=0.3$, and $\Delta _{z}=0.07$.}
\end{figure}

In Fig.2 we present LDOS $\rho (k_{x},k_{y})$ of the top surface for a
semi-infinite sample with $z\in \lbrack 0,-\infty )$. In this case, $k_{x}$
and $k_{y}$ are good quantum numbers. A recursive approach are employed to
calculate the Green's function for a few top layers~\cite{Recursive}. LDOS
at the top surface can be calculated by $\rho _{i}(k_{x},k_{y})=-\frac{1}{%
\pi }$Tr$[\mathrm{Im}G_{ii}(k_{x},k_{y})]$, where $G_{ii}$ is the retarded
green function for the $i$th layer from the top. In the absence of the
Zeeman field, the gapless linear Dirac cone can be observed. Position
dependence of LDOS along the z-axis shows that the states in the Dirac cone
reside dominantly near the top surface, which is characteristic of the
surface states for a topological insulator, and is consistent with the
analytical solution.\cite{Shan-10njp} In the presence of the Zeeman field,
the LDOS shows that a gap of magnitude $2\Delta _{z}$ is opened, as expected.

\begin{figure}[th]
\centering \includegraphics[height=75mm]{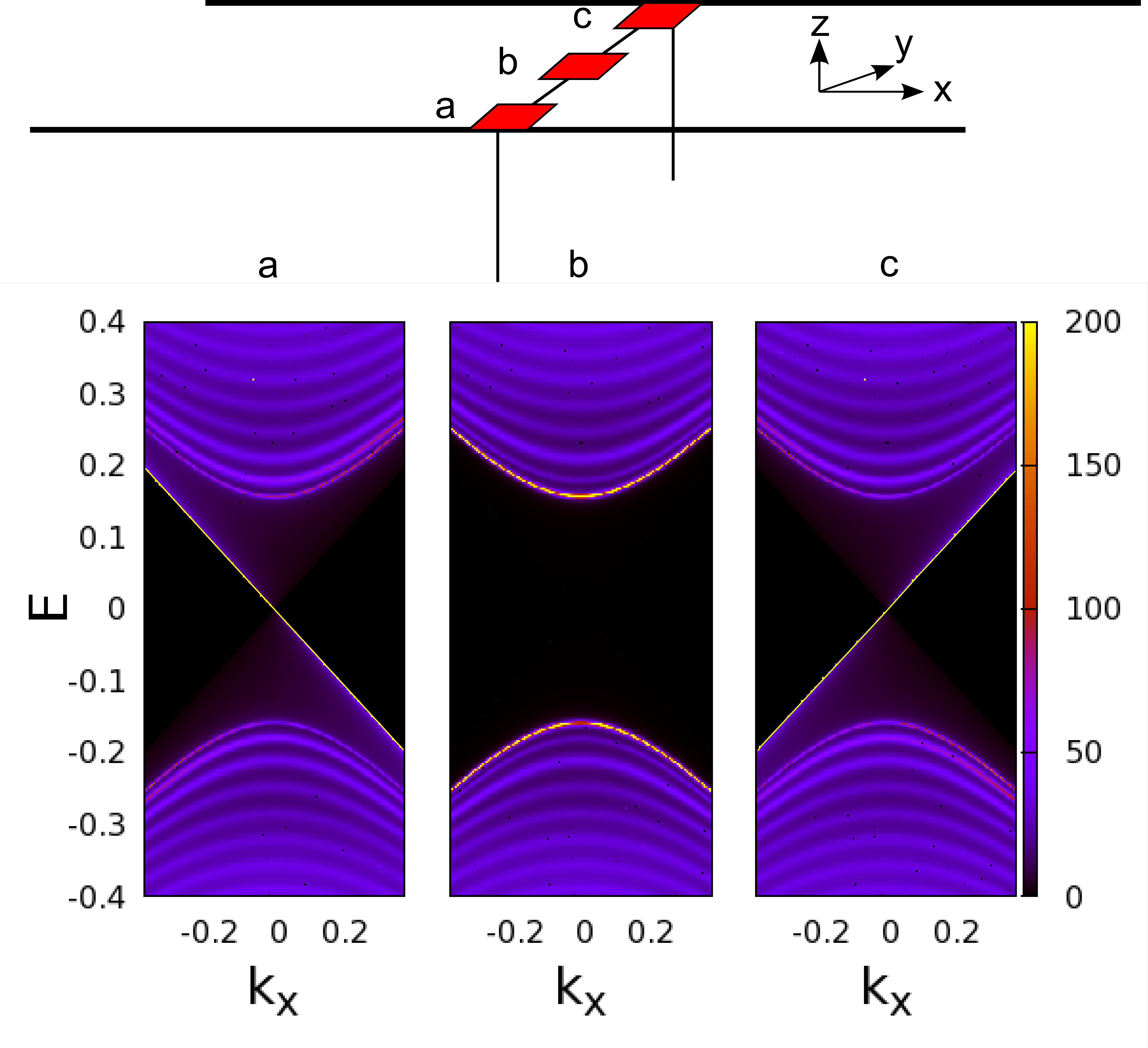}
\caption{(Color online) LDOS on the top surface of a structure that is
infinite in X, finite in Y and semi-infinite in Z direction. Sampling is
taken correspondingly in a, b and c regions as illustrated in the upper
panel.$\Delta _{z}=0.15,M=0.4$, and $L_{y}=30a$ ($a$ is the lattice space)$.$%
}
\end{figure}

To see the distribution of the surface states near the edge of the top
surface, we study a geometry that is finite in the $y$ axis, infinite in the 
$x$ axis and semi-infinite in the $z$ axis, as illustrated in the upper
panel of Fig. 3. In this case only $k_{x}$ is a good quantum number. The
LDOS is $\rho (y,k_{x})=-\frac{1}{\pi }$Tr$[\mathrm{Im}G_{00}(y,k_{x})]$. The
average LDOS at three small different regions (a)--(c) along the $y$%
-direction are shown in the lower panel of Fig.3. It can be seen that the
finite density of states emerges in the gap at the locations (a) and (c)
near the boundaries of the top surface. The finite density of the in-gap
states originates from the penetrated wave functions from the side facets,
which will decay exponentially in space when the energy of the state is
located in the gap of the top surface states. The chiral nature of these
in-gap states can be directly observed: the LDOS at the edge (a) maximizes
at $E=-vk_{x}$ (left going), while the LDOS at the edge (c) maximizes at $%
E=vk_{x}$ (right going). As a result, these states will carry a net chiral
edge current.

We calculate the current density profile of these in-gap states at the fermi
surface $E_{f}$ along the $y$-direction using $\langle J_{x}(y)\rangle
=ie\int_{k_{x}}Tr[v_{x}(r,k_{x})G^{<}(r,k_{x})]$ where $v_{x}=\partial
H/\partial k_{x}$ is the velocity operator and $E_{f}$ is in the Zeeman
splitting gap. In an equilibrium condition we have $%
G^{<}=f(E_{f})(G^{a}-G^{r})$, where $f(E_{f})$ is the Fermi-Dirac
distribution function and we assume zero temperature. In Fig. 4a the current
density distribution along the $y$-direction for the top five layers is
shown. It can be seen that the current is localized dominantly near the
edges in the insulating top layers. To see the distribution of the chiral
edge current in the side facets, we sum up the current density for each
layer over the half the width of $y$: $J_{t}=\sum_{0}^{L_{y}/2}J_{x}(y)$ and
observe its layer dependence. As is shown in Fig. 4b, $J_{t}$ shows the
typical behavior of Friedel oscillation~\cite{Kittel}, which maximize at the
boundary and decays to zero when moving away from the boundary. We see that
the distribution of the chiral edge current is concentrated around the
boundary, even the wave-functions in the metallic region are all extended. 
\begin{figure}[th]
\centering \includegraphics[width=80mm]{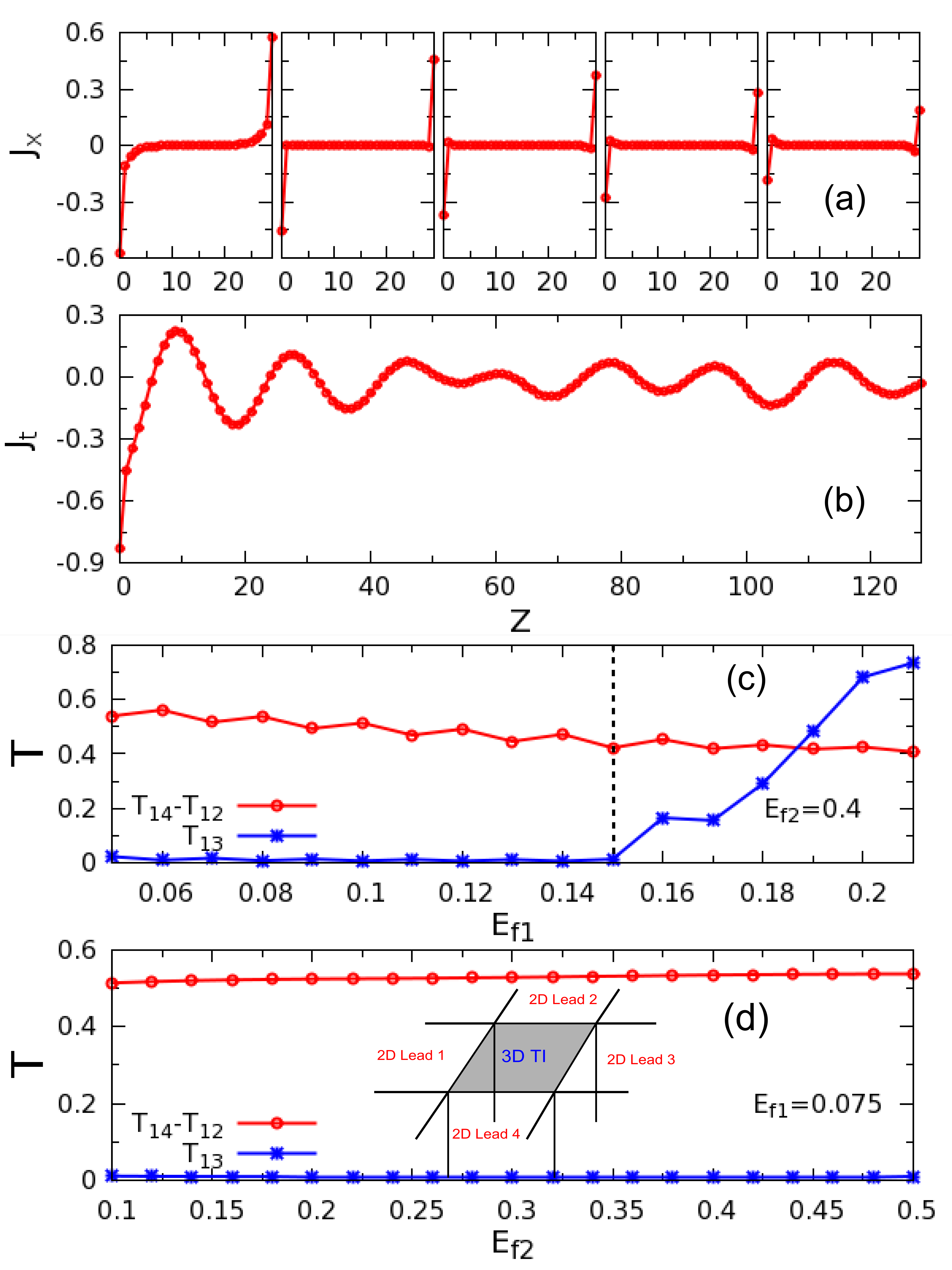}
\caption{(Color online) (a) Local current density $J_{x}$ distribution along
the width of Y ($L_{y}=30a$) for the top 5 layers (the leftest is the 1st);
(b) Total chiral current density $J_{t}$ as a function of layer depth Z; (c)
Transmission coefficients of the 4 terminal device, $E_{f2}$ is fixed, the
dashed line indicates the gap position; (d) same with (c), $E_{f1}$ is
fixed, the insert schematically illustrated the device set up. Other
parameters are $\Delta _{z}=0.15,M=0.4$, $E_{f}=0.075$.}
\end{figure}

After establishing the existence of the chiral surface-edge states, we
calculate the Hall conductance numerically using the Landauer-B\"{u}ttiker
formalism~\cite{Buttiker,Datta,Li-05prb}. The set up of the device is
illustrated in the insert of the Fig. 4d: four 2D metallic leads ($\mu
=1,2,3,4$) are attached to the top surface of a semi-infinite 3D topological
insulator, acting as the measurement electrodes. The Zeeman field is normal
to the top surface. The multi-terminal conductance can be deduced from the
transmission coefficient $T_{pq}$ from the terminal $p$ to terminal $q$, $%
T_{pq}=Tr[\Gamma _{p}G^{R}\Gamma _{q}\Gamma ^{q}]$ where $\Gamma _{p}$ is
determined by the self energy at the terminal $p$~\cite{Datta}. The
calculated transmission coefficients as a function of chemical potentials $%
E_{1f}$ in the 3D topological insulator and $E_{2f}$ in the terminals are
plotted in Figs. 4c and 4d. When the chemical potential $E_{1f}$ is located
in the gap of the surface states, the transmission coefficients show the "half chirality", i.e.,  $T_{pq}-T_{qp}\approx \pm 1/2$ between the two neighboring terminals $p$ and $q$, and $T_{pq} = 0$ between the non-neighboring terminals.  This is different from the 
chirality shown in the usual quantum Hall system, where $T_{pq}-T_{qp} =\pm 1$ between the two neighboring terminals.  It is also important to note that the transmission coefficient
between the two neighboring terminals could be very large due to the presence of the metallic side facets, and it is the difference between the clockwise and anti-clockwise transmission coefficients that shows the "half quantization".

The predicted half quantization can be directly measured.  The straightforward way to observe it in the four-terminal setup is to apply a voltage between the terminals 1 and 3 ($V_{13}$), and measure the current between the terminal 2 and 4 ($I_{24}$).  It is easy to show that the cross-conductance $\sigma_{24, 13}\equiv I_{24} / V_{13} = (e^2/2h)(T_{12} - T{21})$, yielding $e^2/4h$ for the half quantization.   The measurement using the usual six-terminal Hall bar configuration could be more tricky due to the presence of the metallic side facets, which give rise to the finite longitudinal conductance $\sigma_L$.  In the limit of thick sample with $\sigma_L \gg e^2/h$, the Hall conductance $\sigma_H$ should approach to $(4e^2/h) (T_{12} - T_{21})$.  It yields $2e^2/h$ for the half quantization.  It can be distinguished from the usual quantized Hall effect by the finite longitudinal conductance.

To summarize, we find a chiral surface-edge state for a 3D topological
insulator in the presence of Zeeman splitting. Effectively, one can consider
that the original chiral edge mode carrying one conductance quantum $e^{2}/h$
is split into two halves concentrated around the two boundaries between the
insulating domains and metallic region. Its origin is quite different from
those in integer quantum Hall and quantum spin Hall effect, in which there
exist well defined chiral edge states. Here, many conducting channels, each
contributing a fraction of the chirality, may give rise to an effective edge
state concentrated around the domain boundaries, as indicated by the spatial
distribution of the chiral edge current. 
The half quantization can be measured in the multi-terminal
setup.  Moreover, the existence of the effective edge state may be directly observable.
For example, because the spin polarization of Dirac fermions in the surface
states is proportional to the charge current, the spatial oscillatory
pattern of the chiral edge current associating with the effective edge state
could be directly probed by the spatially resolved Kerr rotation technique~%
\cite{Kato}.

The authors thank Z. Fang, Q. F. Sun, W. Yao and F. C. Zhang for helpful
discussions. This work was supported by the Research Grant Council of Hong
Kong under Grant No.: HKU 7037/08P, and HKUST3/CRF/09.

\end{document}